\RequirePackage{fix-cm}
\documentclass[smallcondensed]{svjour3}      
\smartqed  
\usepackage{graphicx}
\usepackage{wrapfig}

 \usepackage{mathptmx}      
\usepackage{latexsym}
\usepackage{fullpage}
\journalname{Hyperfine Interactions}
\begin{document}

\title{VIP 2: Experimental tests of the Pauli Exclusion Principle for electrons}

\author{A. Pichler \and			
				S. Bartalucci \and
				S. Bertolucci \and 
				C. Berucci \and \\
				M. Bragadireanu \and
				M. Cargnelli \and
				A. Clozza \and
				C. Curceanu \and \\
				L. De Paolis \and	
				S. Di Matteo \and
				A. D'Uffizi \and
				J.-P. Egger \and \\
				C. Guaraldo \and
				M. Iliescu \and
				T. Ishiwatari \and
				M. Laubenstein \and \\
				J. Marton \and
				E. Milotti \and
				D. Pietreanu \and
				K. Piscicchia \and
				T. Ponta \and \\
				E. Sbardella \and
				A. Scordo \and
				H. Shi  \and
				D. Sirghi \and
				F. Sirghi \and \\
				L. Sperandio \and
				O. Vazquez-Doce \and
				E. Widmann \and
				J. Zmeskal 
}

\institute{ C. Berucci $\cdot$ M. Cargnelli $\cdot$ T. Ishiwatari $\cdot$ J. Marton $\cdot$ A. Pichler $\cdot$ H. Shi $\cdot$ E. Widmann $\cdot$ J. Zmeskal
					\at
					Stefan-Meyer-Institut f\"ur Subatomare Physik, Boltzmanngasse 3, 1090 Wien, Austria \\
					\email{andreas.pichler@oeaw.ac.at}
					\and
					S. Bartalucci $\cdot$ C. Berucci $\cdot$ M. Bragadireanu $\cdot$ A. Clozza $\cdot$ C. Curceanu $\cdot$ L. De Paolis $\cdot$ A. D'Uffizi $\cdot$ C. Guaraldo $\cdot$ M. Iliescu $\cdot$ D. Pietreanu $\cdot$ K. Piscicchia $\cdot$ T. Ponta $\cdot$ E. Sbardella $\cdot$ A. Scordo $\cdot$ D. Sirghi $\cdot$ F. Sirghi $\cdot$ L. Sperandio  
					\at
					INFN, Laboratori Nazionali di Frascati, C.P. 13, Via E. Fermi 40, I-00044 Frascati(Roma), Italy
					\and
					S. Bertolucci 
					\at
					CERN, CH-1211, Geneva 23, Switzerland
					\and
					M. Bragadireanu $\cdot$ C. Curceanu $\cdot$ D. Pietreanu $\cdot$ D. Sirghi $\cdot$ F. Sirghi
					\at
					IFIN-HH, Institutul National pentru Fizica si Inginerie Nucleara Horia Hulubbei, Reactorului 30, Magurele, Romania
					\and
					S. Di Matteo 
					\at
					Institut de Physique UMR CNRS-UR1 6251, Universit\'e de Rennes, F-35042 Rennes, France
					\and
					J.-P. Egger
					\at
					Institut de Physique, Universit\'{e} de Neuch\^{a}tel, 1 rue A.-L. Breguet, CH-2000 Neuch\^{a}tel, Switzerland
					\and
					M. Laubenstein 
					\at
					INFN, Laboratori Nazionali del Gran Sasso, I-67010 Assergi (AQ), Italy
					\and
					E. Milotti
					\at
					Dipartimento di Fisica, Universit´a di Trieste and INFN-Sezione di Trieste, Via Valerio, 2, I-34127 Trieste, Italy
					\and
					O. Vazquez-Doce
					\at
					Excellence Cluster Universe, Technische Universit\"at M\"unchen, Boltzmannstraße 2, D-85748 Garching, Germany
					\and
					C. Curceanu $\cdot$ K. Pisciccia
					\at
					Museo Storico della Fisica e Centro Studi e Ricerche Enrico Fermi, Piazza del Viminale 1, 00183 Roma, Italy
}

\date{Received: date / Accepted: date} 


\maketitle

\begin{abstract}
Many experiments investigated the violation of the Pauli Exclusion Principle (PEP) since its discovery in 1925. The VIP (VIolation of the Pauli Principle) experiment tested the PEP by measuring the probability for an external electron to be captured and undergo a 2p to 1s transition during its cascading process, where the 1s state is already occupied by two electrons. This transition is forbidden by the Pauli Exclusion Principle. The VIP experiment resulted in a preliminary upper limit for the probability of the violation of the PEP of $4.7 \times 10^{-29}$. Currently a setup for the follow-up experiment VIP 2 is under preparation. The goal of this experiment is to improve the upper limit for the violation of the PEP by two orders of magnitude, by different improvements like enhanced energy resolution of the X-ray detectors and by implementing an active shielding. Here we report currently ongoing performance tests of the new parts of the setup.

\end{abstract}

\section{Introduction}
\label{sec:Introduction}

The austrian physicist Wolfgang Pauli formulated his famous exclusion principle in 1925, which says that two electrons can not be in the same quantum state \cite{Pauli}. For his discovery, he was awarded with the Nobel Prize in 1945. It is one of the fundamental principles in physics, and of the highest importance where many fermion systems are concerned. Therefore, it is important to make precision tests of the PEP. Until now there is no intuitive explanation for the Pauli Exclusion Principle \cite{Feynman}. A possible small violation of the PEP is often quantified in terms of the parameter $\beta$, introduced by Ignatiev and Kuzmin in their (IK) model \cite{IK}. In the IK model, the $\beta$ parameter was introduced by the following relations:\\
\begin{equation}
a^{\dag}\left|0\right\rangle=\left|1\right\rangle, a^{\dag}\left|1\right\rangle=\beta \left|2\right\rangle,a^{\dag}\left|2\right\rangle=0
	\label{eq:IK_beta1}
\end{equation}
\begin{equation}
a\left|0\right\rangle=0, a\left|1\right\rangle=\left|0\right\rangle,a\left|2\right\rangle=\beta \left|1\right\rangle
	\label{eq:IK_beta2}
\end{equation}
Here $a^{\dag},a$ are the ladder operators connecting states with different occupation number $\left|0\right\rangle,\left|1\right\rangle,\left|2\right\rangle$ (forbidden). In an atom, a randomly chosen pair of electrons has the probability of $1-\frac{\beta^{2}}{2}$ to be in a normal antisymmetric state and the probability $\frac{\beta^{2}}{2}$ to be in a non-Paulian symmetric state \cite{TAUP}.

\begin{figure}[htbp]
	\centering
		\includegraphics[width=0.70\textwidth]{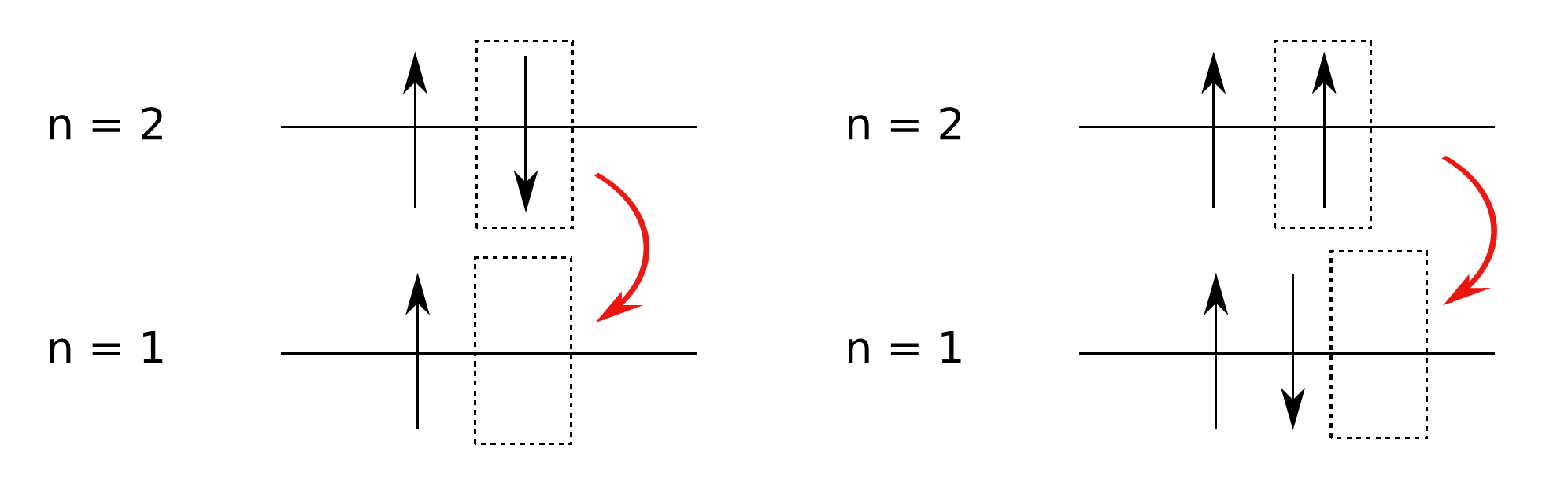}
	\caption{Normal (allowed) 2p - 1s transition with an energy of 8.05 keV for Copper (left) and non-Paulian (forbidden) transition  with an energy of ~7.7 keV for Copper (right).}
	\label{fig:energy_scheme}
\end{figure}

One of the first experiments looking for a small violation of the PEP was conducted by Goldhaber and Scharff-Goldhaber in 1948 \cite{Goldhaber}. It was originally designed to check if the particles that made up beta rays were the same as the electrons in atoms, but it was later used to put an upper bound to the probability of the violation of the Pauli exclusion principle. In this experiment, beta rays were absorbed by a block of lead. The idea of the authors was, that if the 2 kinds of particles were not identical, the beta ray particle could be captured by the atom and cascade down to the ground state without being subject to the PEP. During this cascading process, the X-ray emitted in the 2p to 1s transition would have a slightly different energy than the one from a normal electronic 2p to 1s transition, due to the additional shielding of the 2 electrons which are already in the ground state before the transition happens. This experiment can also be used to validate the PEP, as electrons in a non-Paulian mixed symmetry state also have the possibility to undergo the 2p - 1s transition, with two electrons already in the 1s state prior to the transition. This circumstance is illustrated in figure \ref{fig:energy_scheme} where copper was considered.
This experiment introduced ``fresh'' particles (electrons) to the system. Every time an electron is captured by an atom a new state is formed which has a certain probability to be a mixed symmetry state. Therefore, every capturing process is a test for the PEP. Other experiments however searched for transitions from a normal antisymmetric state to a mixed symmetry state \cite{Reines,Logan}. These transitions are forbidden by the Messiah-Greenberg superselection rule \cite{MG} and therefore the results of those two kinds of experiments cannot be compared. Consequently the concept of introducing ``fresh'' electrons is very important in the search for a violation of the Pauli exclusion principle.

\section{Experimental Method}
\label{sec:ExperimentalMethod}

\begin{figure}[htbp]
	\centering
		\includegraphics[width=0.45\textwidth]{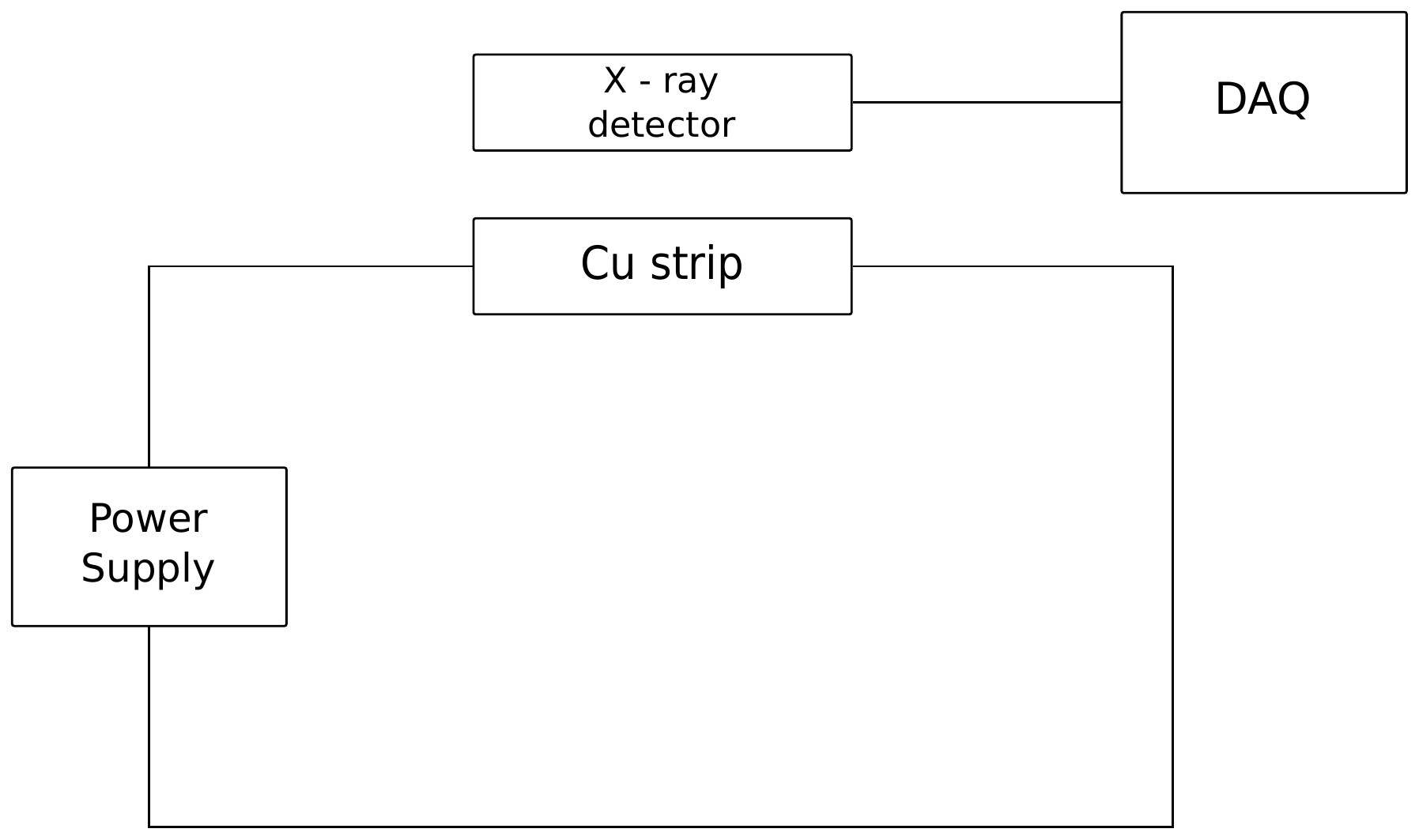}
	\caption{Scheme of the experimental setup for the detection of non-Paulian transitions from Ramberg and Snow \cite{RS}.}
	\label{fig:RS_experiment_scheme}
\end{figure}

The same idea of introducing ``fresh'' electrons (electrons new to the system) to a material to test the PEP was picked up by Ramberg and Snow \cite{RS}. They changed the origin of those electrons from a beta decay source to an electric current. The current of 50 A was applied to a thin copper strip. The X-rays, which resulted from the cascading process of the captured current electrons, were recorded by a proportional tube counter. Figure \ref{fig:RS_experiment_scheme} shows the working principle of the experiment looking for PEP-forbidden atomic transitions.
\subsection{The VIP experiment}
\label{sec:TheVIPExperiment}
The VIP experiment, which started operating in 2006 and took data until 2010, was working similarly as the one conducted by Ramberg and Snow. A current of 40 A was applied to a copper target and the resulting X-ray spectrum was detected. Charged Coupled Devices (CCDs) with an energy resolution of around 320 eV (FWHM) at 8 keV were used as X-ray detectors. Data was taken with and without current. When the current was on, the current electrons had a certain probability to be captured by the target atoms, to form a mixed symmetry state and to undergo the non-Paulian 2p - 1s transition. Without current, this process was not possible, and the resulting spectrum was recorded to get the number of background events.
The setup was tested at LNF-INFN, where measurements were performed from 21$^{st}$ November - 13$^{th}$ December 2005. The data analysis followed the arguments of Ramberg and Snow \cite{RS} and led to a probability that PEP is violated of \cite{VIP2006}:
\begin{equation}
	\frac{\beta^{2}}{2}\leq4.5\times10^{-28}
\end{equation}
In order to diminish the background, which was mainly caused by cosmic rays and natural radioactivity, the experiment was later located in the low-background environment of the underground laboratory LNGS-INFN. The experiment took data from spring 2006 until 2010. The analysis of the data resulted in a preliminary upper limit for the probability for a violation of the Pauli exclusion principle of \cite{Catalina1,Catalina2}:
\begin{equation}
	\frac{\beta^{2}}{2}\leq4.7\times10^{-29}
\end{equation}
The result is expressed in terms of the parameter $\beta$ mentioned before.

\subsection{From VIP to VIP 2}
\label{sec:FromVIPToVIP2}

Currently, we are working on the follow-up experiment VIP 2. One of the major changes will be the use of Silicon Drift Detectors (SDD) instead of CCDs as X-ray detectors, which have an even better energy resolution than CCDs and most important, the detectors have a timing resolution (FWHM) of around 400 ns \cite{TAUP}. This enables the identification of background events and consequently an active shielding. For this purpose, scintillator bars read out by Silicon photomultipliers are installed around the target and the SDDs. They are used to identify those events in the SDDs which are in coincidence with an external event in the scintillator, and are therefore probably triggered by background. Another improvement is the implementation of a new target, to which a current of at least 100 A can be applied. A summary of the changes is given in table \ref{tab:VIP2_improvement}. These improvements are essential for reaching the goal of reducing the upper limit for the probability of a violation of the PEP to the order of $10^{-31}$. 
\begin{table}
\centering
\caption{List of improvements of VIP 2 compared to VIP \cite{VIP_proposal}.}
\label{tab:VIP2_improvement}
	
		\includegraphics[width=0.55\textwidth]{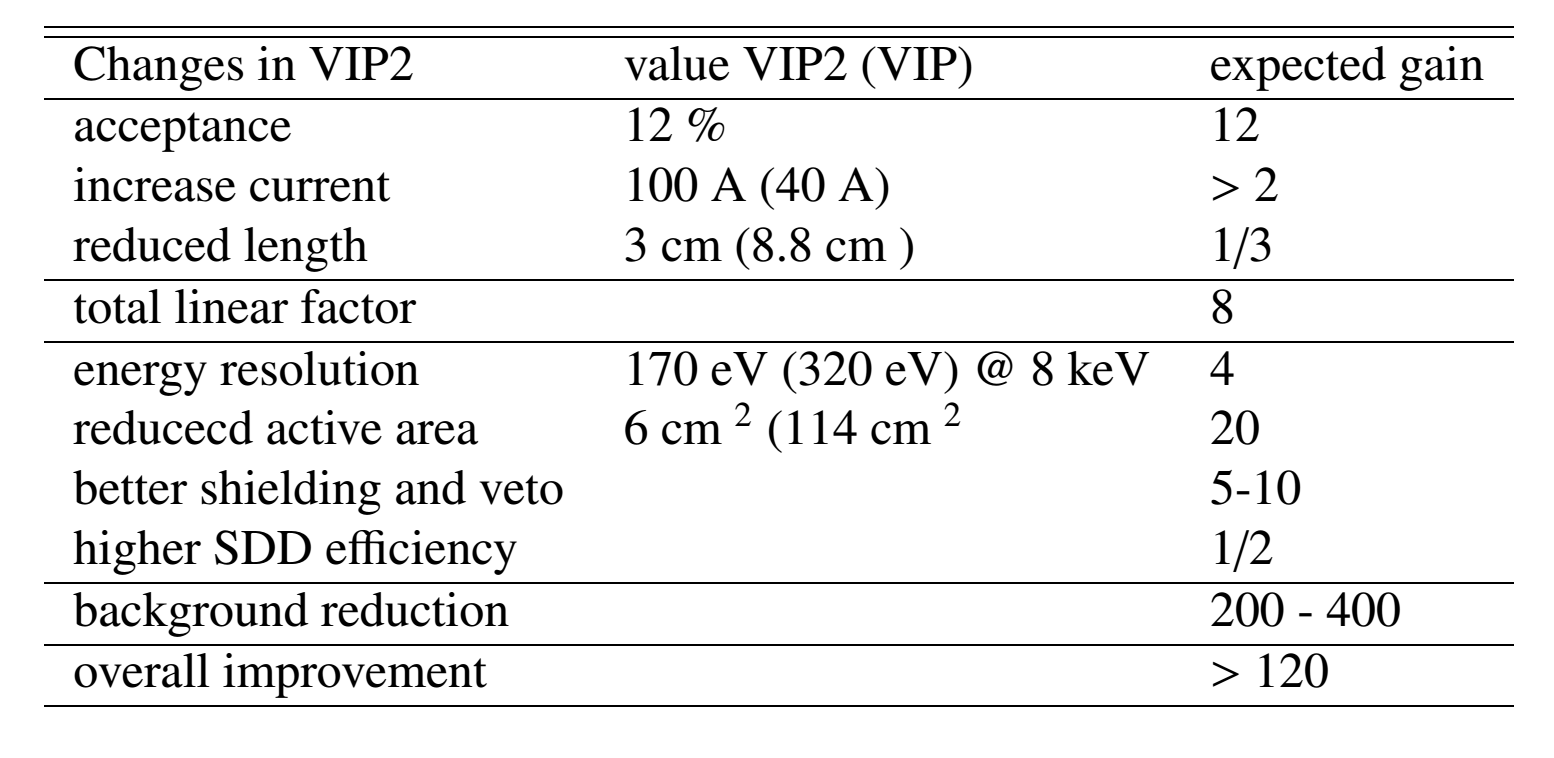}

\end{table} 

\newpage

\subsection{SDD energy resolution}
\label{sec:SDDEnergyResolution}

An important point for the improvement of the PEP violation limit is the energy resolution of the used X-ray detectors.
\begin{wrapfigure}{r}{0.4\textwidth}
	\centering
		\includegraphics[width=0.35\textwidth]{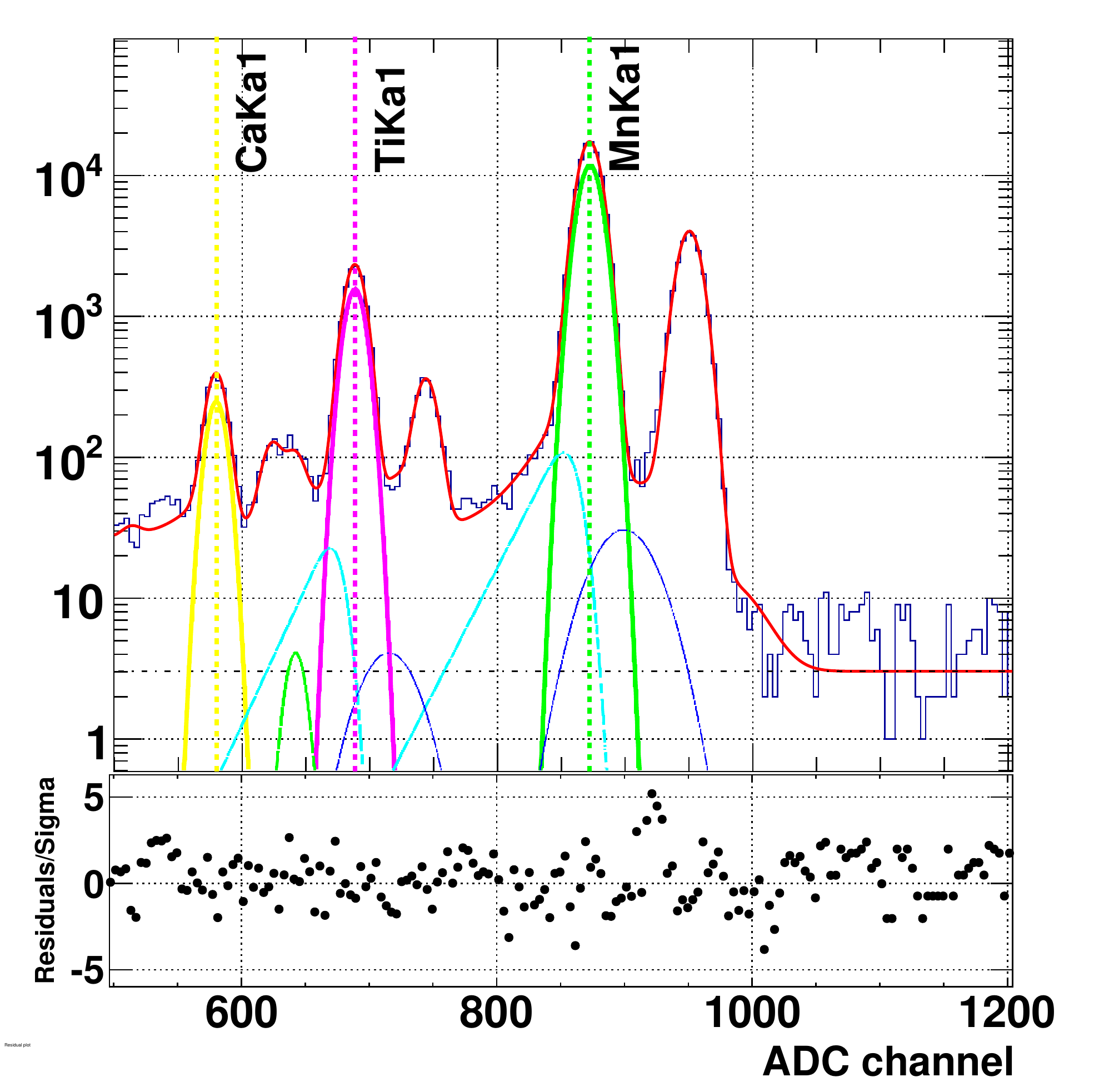}
		\caption{ADC spectrum recorded by one SDD with a $^{55}Fe$ source and a Ti calibration foil. We used the Ti and Mn K$\alpha_{1}$ lines to determine the energy scale.}
	\label{fig:20140821adc14calib}
\end{wrapfigure}
For testing an $^{55}Fe$ source irradiated a 5 $\mu$m titanium foil and 2 arrays of Silicon Drift Detectors. Each of these arrays was made of 3 detectors with around 1 cm$^{2}$ active area. This radiation was not only registered by the detectors, but also caused excitations and subsequent emission of K$\alpha$ and K$\beta$ fluorescence photons from the titanium foil. These photons were also recorded.

The SDDs were connected to an argon cooling cycle with a cooling finger. Their temperature was stable during the measurement at around 125 K. The $^{55}Fe$ source was put onto the evacuated setup which had the SDDs and the titanium foil inside, while still being able to irradiate them through a window made of kapton. The results for one detector are shown in figure \ref{fig:20140821adc14calib}. The function we used to fit the SDD spectra is discussed in detail in \cite{Siddharta} by the SIDDHARTA group, where the same detectors were used. Using this fit function it was possible to get small residuals, which are shown in figure \ref{fig:20140821adc14calib}. From the shape and the position of the K$\alpha$ lines we get a typical energy resolution of around 150 eV (FWHM), thus fulfilling our requirements.


The copper target was cooled by flowing water through a cooling line next to it. The current was increased starting from 80 A up to 180 A. Through the water cooling the target temperature was stabilized at around 20 $^{\circ}$C. This stability at high current is important for VIP 2, as an increase in current improves the outcome of the experiment linearly.

\section{Outlook}
\label{sec:Outlook}

In the near future the complete active shielding with scintillators and the SDDs will be implemented. Afterwards data taking will start in the laboratory at Stefan Meyer Institute in Vienna. We will test the rejection efficiency of the active shielding, the stability of the SDDs, the cryogenic system and the water cooling system of the target. Then the setup will be transported to the underground laboratory of Gran Sasso for the long term (3-4 years) data taking. Finally we expect to improve the upper limit of PEP for electrons by at least 2 orders of magnitude.

\begin{acknowledgements}

We want to thank H. Schneider, L. Stohwasser and D. St\"uckler from the Stefan Meyer Institute for their important contributions for the design and the construction of the VIP 2 setup and the staff of the INFN-LNGS laboratory for the support during all phases of the experiment. We acknowledge the support from the: HadronPhysics FP6 (506078), HadronPhysics2 FP7 (227431), HadronPhysics3 (283286) projects, EU COST Action 1006 (Fundamental Problems in Quantum Physics), Austrian Science Fund (FWF), which supports the VIP 2 project with the grant P25529-N20 and Centro Fermi (project: Open problems in quantum mechanics).

\end{acknowledgements}

\end{document}